\documentclass[aps,prl,twocolumn,superscriptaddress]{revtex4}
\usepackage{graphicx}
\usepackage{xcolor}
\input epsf
\usepackage{amsmath,amssymb}
\begin{document}
\title{Magnetization Reversal Across Multiple Serial Barriers in a Single Fe$_3$O$_4$ Nanoparticle}
\author{Sagar Paul}
\affiliation{Department of Physics, Indian Institute of Technology Kanpur, Kanpur 208016, India}
\author{Ganesh Kotagiri}
\affiliation{Department of Physics, Indian Institute of Technology Kanpur, Kanpur 208016, India}
\author{Rini Ganguly}
\affiliation{\mbox{Univ.} Grenoble Alpes, CNRS, Grenoble INP, Institut N\'eel, 38000 Grenoble, France}
\author{Annapoorni Subramanian}
\affiliation{University of Delhi, Delhi 110021, India}
\author{Herv\'{e} Courtois}
\affiliation{\mbox{Univ.} Grenoble Alpes, CNRS, Grenoble INP, Institut N\'eel, 38000 Grenoble, France}
\author{Clemens B. Winkelmann}
\affiliation{\mbox{Univ.} Grenoble Alpes, CNRS, Grenoble INP, Institut N\'eel, 38000 Grenoble, France}
\author{Anjan K. Gupta}
\affiliation{Department of Physics, Indian Institute of Technology Kanpur, Kanpur 208016, India}
\date{\today}

\begin{abstract}
Depinning of nanoscale magnetic textures, such as domain walls, vortices and skyrmions, is of paramount importance for magnetic storage and information processing. We measure time-resolved magnetic switching statistics of an individual, non-single-domain Fe$_3$O$_4$ nanoparticle using a micrometer-scale superconducting quantum interference device. Surprisingly, a strong narrowing of the waiting-time distributions before reaching the final state is observed as compared to the exponential distribution expected for a single barrier. The magnetization reversal across the nanostructure is thus shown to result from multiple serial barriers in the minimum energy pathway.
\end{abstract}

\maketitle
Ferromagnetic nanostructures display intriguing physics and applications in wide areas, including digital memory \cite{domain-wall-racetrack,domain-wall-logic}, information processing \cite{Sky-elect,Sky-qbit,Sky-bimeron-1} and biomedicine \cite{bio-medicine-1}. In the smallest ferromagnetic structures, exhibiting a single magnetic domain, the magnetisation reversal occurs through a coherent rotation described by the Stoner-Wohlfarth model \cite{stoner-wohlfarth,Co-mnp-werns}. In such systems, the switching time distributions, close to the switching field, are well described by a single exponential with a characteristic time $\tau$.
%, and the switching field histograms are narrow and well defined.
The temperature dependence of $\tau$ is well described by the N\'eel-Brown model \cite{Neel-Brown-1,Neel-Brown-1a,Neel-Brown-2,Neel-Brown-3}, based on thermal activation, at high temperatures, and quantum tunneling at low temperatures \cite{Wernsdorfer-Ni-nanowire,Wernsdorfer-NeelBrown}. With increasing size of the nanostructures, the magnetisation reversals occur principally through curling mode involving inhomogeneous rotation of spins or through propagation and annihilation of a vortex or a domain wall.
In a nanowire, the reversal occurs through vortex-pair nucleation and annihilation and the switching time distributions show a stretched exponential decay, with wide switching field histograms. This is attributed to the presence of many parallel minimum energy pathways (MEPs), with different energy barriers, for the vortex.

In a single nanoparticle, from micro-magnetic simulations based on a finite-element solution of the Landau-Lifshitz-Gilbert equation, it is expected that the reversal occurs via the following main steps \cite{suppl-info}. When a threshold applied magnetic field is reached, a curling mode or a vortex nucleates at the nanoparticle surface, a process during which only a fraction of the total spins undergo reversal. As the field is further increased, this vortex or curling mode traverses through the nanoparticle volume, by following one of several possible paths. Eventually, at the annihilation field, the magnetization reversal completes, barring some surface spins, as the vortex annihilates. This reversal process can be inferred to some extent by the analysis of the magnetization cycle. Nevertheless, understanding the mechanisms at stake during the vortex traversal of the nanoparticle requires more information.

In this Letter, we present magnetic switching-field and switching-time histograms of single ${\rm Fe}_3{\rm O}_4$ nanoparticles. Both distributions are found to be way too narrow, and thus too deterministic, to be compatible with a magnetization reversal limited by a single energy barrier. In contrast, we show that a scenario involving several barriers in series accounts well for our observations. In the experiments, the number of serial barriers is found to reduce as the switching field is approached. Eventually, very close to the thermodynamic switching field, the single barrier limit is recovered.

Micron or nanometer scale superconducting quantum interference devices ($\mu$- or nano-SQUIDs) have been the most successful probe till date for magnetization reversal studies on individual magnetic nanoparticles \cite{Co-mnp-werns, Wernsdorfer-NeelBrown, hightc-1} and nanowires \cite{Wernsdorfer-Ni-nanowire, hightc-2, hightc-3} by direct coupling to the SQUID loop. Here we study the magnetism of individual $\rm Fe_3O_4$ nanoparticles of diameter 150 $\pm$ 20 nm [$F\#1-3$] using Nb $\mu$-SQUIDs working in a non-hysteretic regime obtained by an optimum shunt \cite{Sourav-inductiveshunt}. This measurement setup including a 3D vector magnet has been used earlier for probing magnetic anisotropy in permalloy nanowires \cite{Sagar-jmmm}. The $\mu$-SQUID's sensitivity was further improved by using a low temperature SQUID-array amplifier. More experimental details on the measurements, nanoparticle synthesis and its placement on the $\mu$-SQUID are given in \mbox{Suppl. Info.} \cite{suppl-info}.

\begin{figure}[t!]
	\centering
	\includegraphics[width=\columnwidth]{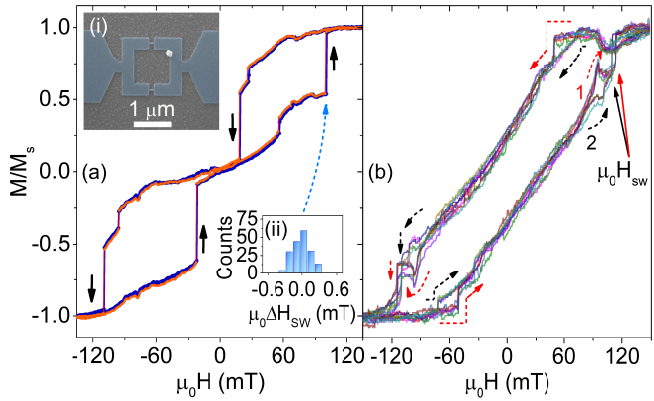}
	\caption{(a) Measured $M-H$ loops of the $\rm Fe_3O_4$ nano-particle $F\#1$ for a field at an in-plane angle $\theta$ = - 45$^\circ$ and at $T$ = 4.0 K. The loop has two major jumps (marked by arrows) in each field sweep direction due to vortex nucleation and annihilation. Inset (i) shows the electron micrograph of the $\mu$-SQUID (in false color) with the measured $\rm Fe_3O_4$ nano-particle. Inset (ii) shows the switching field histograms at the vortex annihilation at positive field. (b) Measured $M-H$ loops for the Fe$_3$O$_4$ nano-particle $F\#2$ (at T = 4.2 K, $\theta$ = 0$^0$) exhibiting two parallel paths marked by red and black arrows.}
	\label{fig1:fe3o4-M-H}
\end{figure}

Figure \ref{fig1:fe3o4-M-H}(a) shows the measured magnetization versus field (M-H) curves of a $\rm Fe_3O_4$ nanoparticle sample $F\#1$, exhibiting hysteresis and two prominent jumps in each path. Each sweep shows two main jumps corresponding to a vortex nucleation and annihilation. When the magnetic field $H$ is swept back and forth repeatedly, the annihilation or nucleation does not always occur precisely at the same field value, leading to a distribution in measured $H_{\rm sw}$ as shown in \mbox{Fig.} \ref{fig1:fe3o4-M-H}(a) inset histogram. In contrast, another nano-particle named $F\#2$ displays a non-zero remanence, see \mbox{Fig.} \ref{fig1:fe3o4-M-H}(b), suggesting a curling mode. Each jump in a sweep occurs at one of two distinct field values, leading to observation of two disjoint peaks in the switching field histograms. The presence of two distinct reversal pathways is inferred from the correlation between the nucleation and annihilation field values. Two paths can arise from two slightly different trajectories of the vortex. Another possibility is the vortex chirality in the sense of a vortex with clockwise or anticlockwise spin order. A defect can possibly lead to an affinity for one chirality vortex over the other leading to difference in nucleation and annihilation fields. It may not always be possible to differentiate between parallel pathways, particularly if the switching field distributions of individual pathways have a width exceeding their separation. In the present study, we observe only one or at most two pathways or MEPs, each displaying a very narrow $H_{\rm sw}$ distribution.

\begin{figure}[t!]
	\centering
	\includegraphics[width=1\columnwidth]{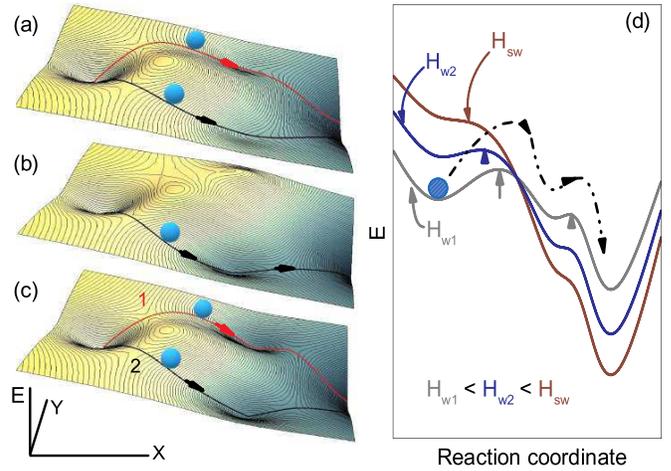}
	\caption{(a), (b) and (c) show, respectively, the schematic of the free energy landscape for fields close to the switching field to illustrate two parallel minimum energy paths, two serial barriers in a single path and two parallel paths with two serial barriers in each. (d) shows the schematic of the energy landscape for two serial barriers at different applied fields close to the switching field $H_{\rm sw}$. This illustrates how the barriers disappear one by one as one tilts the energy landscape by increasing $H$.}
	\label{fig2:schematic}
\end{figure}

In a magnetic nanostructure, a vortex or a vortex pair is expected to follow an MEP in an energy landscape \cite{MEP-FM-1,MEP-FM-3} determined by the external field, the crystalline anisotropy, the exchange energy, the demagnetizing field and defects. The switching time statistics can provide a way to probe the nature of the MEP(s) at any fixed field close to the threshold $H_{\rm sw}$. Multiple parallel MEPs can arise for a vortex, see \mbox{Fig.} \ref{fig2:schematic}(a), with each path exhibiting different barriers, see \mbox{Fig.} \ref{fig2:schematic}(b,c). Along an MEP, the escape rate from an energy minimum depends on the product of the attempt rate, determined by the dynamics near the minimum, and the probability of overcoming the barrier by thermal activation or quantum tunneling.

We define the cumulative distribution function (CDF) for \emph{not switching} during a time $t$ as $P(t)$, as well as the probability density function (PDF) $p(t)$ with $p(t)dt$ being the probability of switching during the time interval $t$ to $t+dt$. These two are related by $P(t)=1-\int_0^tp(s)ds$. The PDF provides the histogram of the waiting times before switching at a fixed waiting field $H_w$ close to the switching field $H_{\rm sw}$. For the case of a single barrier, the probability of switching in a time interval $dt$ is $dt/\tau$ with $\tau$ as the mean switching time. With this, one gets $p(t)=\tau^{-1}\exp(-t/\tau)$, and $P(t)=\exp(-t/\tau)$.

The case of $N$ independent parallel paths, see \mbox{Fig.} \ref{fig2:schematic}(a), is a relevant scenario to analyze. Here, the $i^{\rm th}$ path has a single barrier, with an associated transition rate $\tau_i^{-1}$. This leads to a CDF
\begin{equation}
P_{\rm par}(t) = \sum_{i=1}^{N} w_ie^{-t/\tau_i}.
\label{eq:Parallel-Barriers}
\end{equation}
Here $w_i$ the probability of selection of the $i^{\rm th}$ path with $\sum_i w_i=1$. The effective mean transition time is then $\tau_{\rm eff}=\sum_i w_i\tau_i$. Many parallel barriers can lead to a behavior close to the stretched-exponential relaxation given by $P_{\rm str}(t) = \exp[-(t/\tau)^\beta]$
with $\beta<1$. A precise stretched-exponential results from a systematic probability distribution of transition rates \cite{stretched-expo-from-distribution}. In the case of a statistical ensemble of particles, this could be easily justified but for an isolated particle with only a few parallel paths there is no reason \emph{a priori} to expect the same.

For $N$ barriers of equal transition rate $\tau^{-1}$ in series, the CDF writes \cite{suppl-info}
\begin{equation}
P_{Ne}(t)=e^{-Nt/\tau_{\rm eff}}\sum_{k=0}^{N-1}\frac{(Nt/\tau_{\rm eff})^k}{k !},
\label{eq:N-equal-barriers}
\end{equation}
where $\tau_{\rm eff}=N\tau$ is the overall mean switching time. Note that this expression is the product of $e^{-Nt/\tau_{\rm eff}}$ and the truncated polynomial expansion of $e^{+Nt/\tau_{\rm eff}}$. In the case of a set of $N$ {\em unequal} values of $\tau_i$, the CDF is
\begin{align}
P_{Nu}(t)=\sum_i\frac{\tau_i^{N-1}}{f_i(\tau_1,\tau_2...\tau_N)}e^{-t/\tau_i},
\label{eq:N-unequal-barriers}
\end{align}
with $f_i(\tau_1,\tau_2...\tau_N)=\prod_{j\neq i}(\tau_i-\tau_j)$. The transition of $P_{Ne}(t)$ from 1 to 0 becomes steeper with increasing $N$ \cite{suppl-info} and the corresponding PDF $p(t)$ becomes a more sharply peaked function of width $\propto \tau_{\rm eff}/\sqrt{N}$. The CDF thus spreads over a much narrower time window, as compared to an exponential relaxation. Also the distribution of $H_{\rm sw}$ is narrower, though the details of its histogram will depend on the dependence of $\tau_i$ on the applied field and its sweep rate. For a single field-dependent barrier, the probability distribution of $H_{\rm sw}$ has been discussed by Kurkij$\ddot{\rm a}$rvi \cite{kurki}.

\begin{figure}[t!]
	\centering
	\includegraphics[width=1\columnwidth]{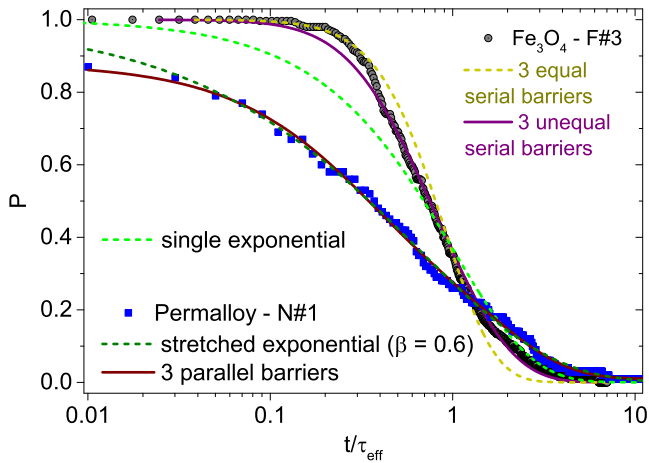}
	\caption{A comparison of experimental CDF of not switching for Fe$_3$O$_4$ nanoparticle at 129.2 mT and that of a permalloy nanowire at 63 mT. The horizontal time axis is in units of respective $\tau_{\rm eff}$. The fits to three equal ($\tau/3=\tau_{\rm eff}$ = 25.73 s) and unequal serial barriers (see Table \ref{table1:fits}) are shown for the nanoparticle data. The fits to the stretched exponential ($\beta$ = 0.6 and $\tau=0.65 \times \tau_{eff}$ = 3.23 s) and three parallel barriers (see Table 2 in \mbox{Suppl. Info.} \cite{suppl-info}) are shown for the nanowire data. The green line shows an exponential with average switching time as $\tau_{\rm eff}$.}
	\label{fig3:model}
\end{figure}

In order to illustrate the immense difference of behaviors that can be observed in various magnetic micro or nano-structures, \mbox{Fig.} \ref{fig3:model} shows the measured probabilities of not switching $P(t)$ for a permalloy nanowire and a ${\rm Fe}_3{\rm O}_4$ nanoparticle. In units of average switching-time $\tau_{\rm eff}$, the permalloy nanowire data spreads over more than three time decades. Moreover, it fits to the multiple parallel barriers model or to a stretched exponential \cite{suppl-info}. This is similar to the Ni nanowire studied by Wernsdorfer et al \cite{Wernsdorfer-Ni-nanowire}. On the other hand, the ${\rm Fe}_3{\rm O}_4$ nanoparticle data only have about a decade spread. A similar behavior has been reported in amorphous Co particles with compressed-exponential fits \cite{werns-comp-exp}. The reduced spread of the CDF in ${\rm Fe}_3{\rm O}_4$ nanoparticle is the main topic of the following of this Letter.

\begin{figure}[t!]
	\centering
	\includegraphics[width=1\columnwidth]{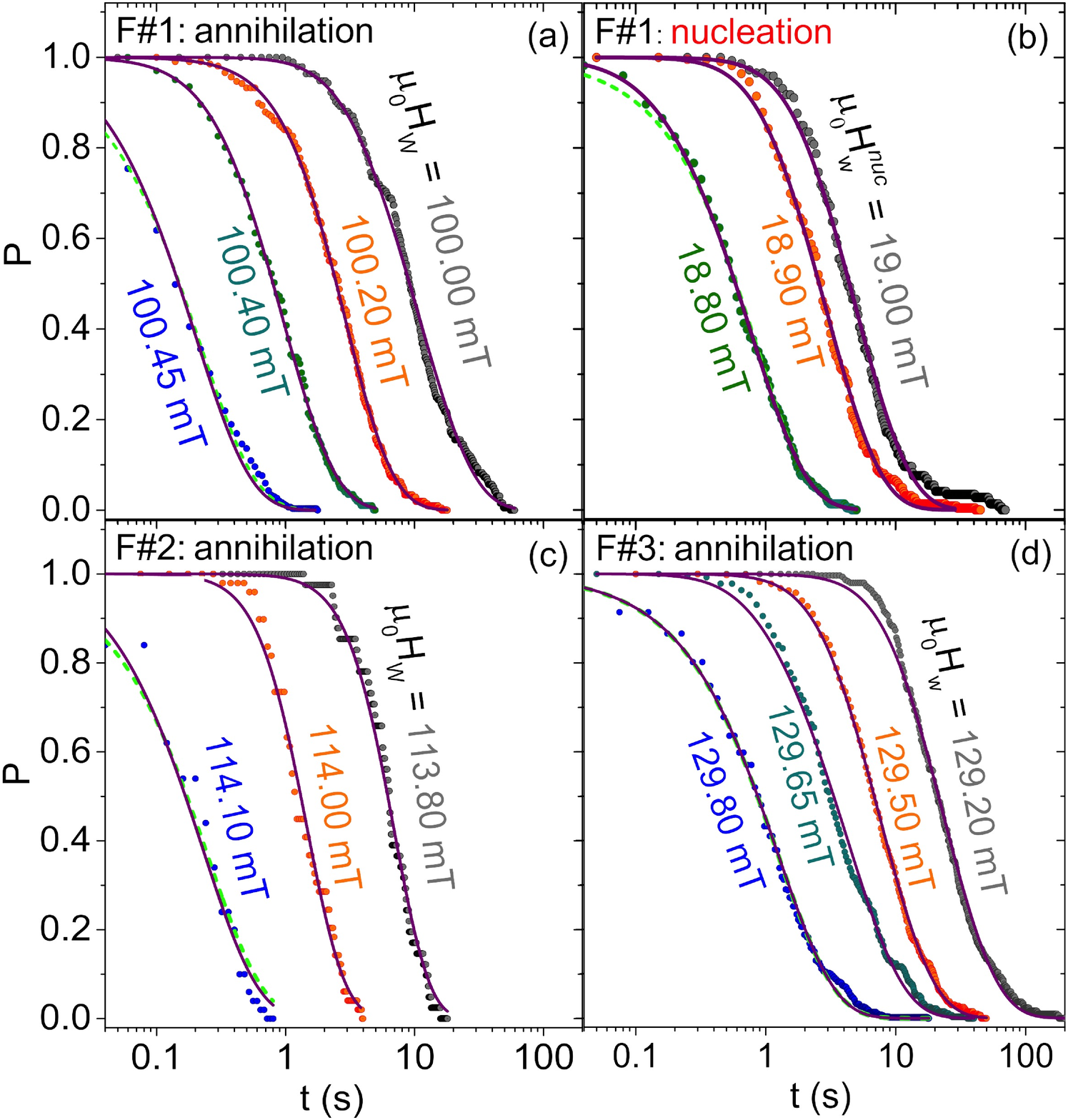}
	\caption{Experimental CDF of not switching vs time (dots) for $\rm Fe_3O_4$ nano-particle $F\#1$ at $\theta$ = - 45$^0$, $T$ = 4.0 K obtained at waiting fields close to (a) annihilation and (b) nucleation. (c) and (d) show the CDF obtained for $F\#2$ at T = 4.2 K, $\theta$ = 0$^0$ and $F\#3$ at T = 2.0 K, $\theta$ = 60$^0$ respectively. The purple lines are fits to the three unequal barrier model and the green dashed lines are fits to single exponential.}
	\label{fig5:Fe3O4-CDF}
\end{figure}

From micromagnetic simulations, it is seen that in a nanomagnet of size below 200 nm a nucleated vortex needs to cross a threshold position in order to annihilate and complete the magnetization reversal. Thus, in the absence of defects one can expect the energy landscape to exhibit two relatively deep and well-separated minima that determine the vortex position near nucleation and near annihilation. This will result in a single barrier along the MEP. However, in presence of defects, the energy landscape may also exhibit other bulges, and dents in which the vortex can get trapped, thus increasing the effective number of barriers. We qualitatively interpret the observed CDF of ${\rm Fe}_3{\rm O}_4$ nanoparticle as due to the presence of multiple barriers along the MEP for magnetization reversal.

\begin{figure*}[t!]
	\centering
	\includegraphics[width=0.75\textwidth]{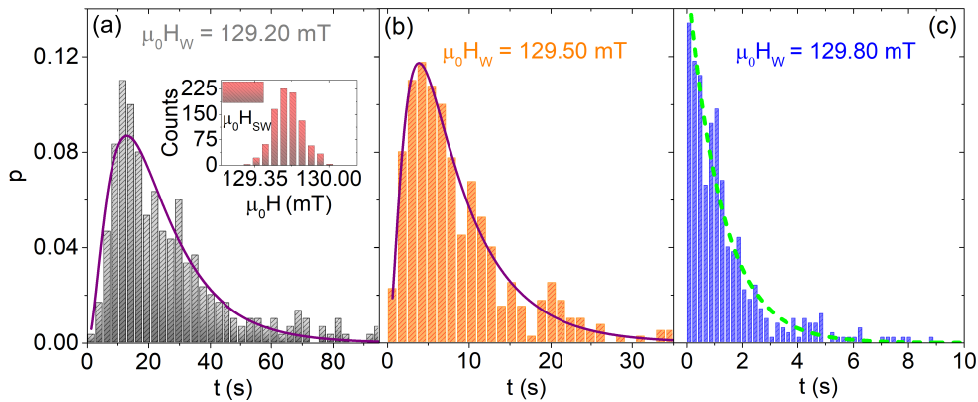}
	\caption{Waiting time histograms for $\rm Fe_3O_4$ nano-particle $F\#3$ at different waiting fields $H_w$, T = 2.0 K and $\theta$ = 60$^0$. A large number ($\sim$ 400) of switching data was acquired. The purple line in (a), (b) shows a fit to the PDF corresponding to a unequal barriers model while the green dashed line in (c) shows a fit to an exponential PDF. The inset is a switching field histogram for this particle with 800 counts obtained at the same temperature, angle and at a sweep rate of 0.3 mT/s.}
	\label{fig6:Fe3O4-pdf}
\end{figure*}

In order to have a more quantitative approach, let us consider the switching statistics in the three studied nanoparticles and at various in-plane angles of the magnetic field. In $F\#1$ and $F\#3$, only one reversal path was seen in $M-H$ loops for every studied angle, temperature and sweep rate. For $F\#2$, we used the early signature of bifurcation (during nucleation) to select and probe the switching time statistics in each path. Figure \ref{fig5:Fe3O4-CDF} shows the CDF for different waiting field $H_w$ values near annihilation and nucleation for the three studied $\rm {\rm Fe}_3{\rm O}_4$ nano-particles. This data could not be fitted with a two-serial-barrier model, nor to any parallel barrier model. In contrast, a nice fit is obtained with a three-serial-barrier model for every data set, as shown by the full lines in \mbox{Fig.} \ref{fig5:Fe3O4-CDF} panels. As seen from the reduced $\chi$-square $\chi_{\rm r}^2$ values, the agreement is remarkable.

\begin{table}[b!]
	\centering
		 \begin{tabular}{|p{1cm}|p{1.1cm}|p{1cm}|p{1cm}|p{1cm}|p{1cm}|p{1.1cm}|}
		\hline
		Dev.& $\mu_0H_w$&\multicolumn{5}{|c|}{3 Unequal Serial Barriers}\\
		no.&(mT)&$\tau_1(s)$&$\tau_2$(s)&$\tau_3(s)$&$\tau_{\rm eff}(s)$&$\chi_r^2\times10^4$ \\
		\hline
		%\hline
		F$\#$1&100.00  &10.79&0.82&0.82&12.43&4.5 \\
		&100.20 & 2.32&0.81&0.00&3.13&0.7\\
		&100.40 &0.89&0.17&0.00&1.06&1.2\\
		&100.45 &0.20&0.01&0.00&0.21&4.7\\
		\hline
		F$\#$3&129.20  & 21.56 & 3.00 & 3.00 &27.56&2.1\\
		&129.50  & 7.06 & 0.91 & 0.91 &8.88&0.8\\
		&129.65  & 4.24 & 0.19 & 0.19&4.62&2.6\\
		&129.80 & 1.25 & 0.00 & 0.00 &1.25&1.3\\
		\hline
	\end{tabular}
\caption{\label{table1:fits}Fitting parameters for CDF for not switching for different $H_w$ for Fe$_3$O$_4$ devices corresponding to \mbox{Fig.} \ref{fig5:Fe3O4-CDF}. Here, $\tau_{\rm eff}$ is the effective mean switching time. Note that $\tau_{\rm eff}$ decreases rapidly and $\tau_{2,3}$ decreases to zero with increasing field.}
\end{table}

A systematic evolution of the fitting parameters of the three-serial-barrier model with increasing $H_{\rm w}$ is presented in Table \ref{table1:fits}. The mean switching time $\tau_{\rm eff}$ decreases rapidly with increasing $H_{\rm w}$ and two out of the three times, \mbox{i.e.} $\tau_{2,3}$, gradually decrease to zero with increasing $H_{\rm w}$. This indicates that two of the three serial barriers disappear as the switching field is approached. This aspect is better seen in the histograms, corresponding to the PDF, in \mbox{Fig.} \ref{fig6:Fe3O4-pdf} for $F\#3$. Remarkably, for the waiting field closest to the thermodynamic switching field, the PDF of a single exponential appears to be the only choice. This is markedly different from the two other histograms in the sense that it does not show a decline down to the smallest waiting times. The disappearance of intermediate barriers when the magnetic field is close to the thermodynamic switching field is fully consistent with the picture of \mbox{Fig.} \ref{fig2:schematic}(d). When the magnetic field is increased, the whole potential profile is tilted, see \mbox{Fig.} \ref{fig2:schematic}(d). Some dents will cease to be actual energy minima, thus reducing the number of barriers along the MEP.

A comparison of these data can also be made with a compressed exponential and a log-normal distribution \cite{suppl-info}. The latter gives a good agreement. A log-normal distribution can indeed be used for describing multiplicative processes \cite{shockley-log-norm,montroll-log-norm,Imry-log-norm} where serial barriers are crossed simultaneously rather than one by one.

In conclusion, while the various studied Fe$_3$O$_4$ particles differ in their detailed M-H loop, thus suggesting that the reversals happen through a vortex or through a curling mode, all of them show strikingly narrow switching field histograms. At a given applied field close to the switching field, the relaxation to the ground state is best described by a model involving few barriers in series. The number of barriers required to fit the data reduces as the waiting field is increased, down to one very close to the thermodynamic switching field. For a system with serial barriers, obtaining experimentally the switching time histograms is much easier, due to much narrower spread, than in a system with parallel barriers. On the same note, such a sharp and definite switching in a non single domain particle draws attention towards the applications in decisive switching. The present serial-barrier model can also apply to other multi-state systems. In particular, it may help in understanding the role of defects in manipulating the topological magnetic textures such as skyrmion and domain walls in racetrack-type memory devices.

This work is supported by project 5804-2 from CEFIPRA, SERB-DST of the Government of India, ANR contract Optofluxonics 17-CE30-0018 and LabEx LANEF (ANR-10-LABX-51-01) project UHV-NEQ. WE also acknowledge L. Buda-Prejbeanu for interesting discussions.

\end{document}